# Functionalized Thallium Antimony Films as Excellent Candidates for Large-Gap Quantum Spin Hall Insulator

Run-wu Zhang,[a] Chang-wen Zhang*,[a] Wei-xiao Ji,[a] Sheng-shi Li,[b] Shi-shen Yan,[b] Ping Li,[a] and Pei-ji Wang,[a]

[a] School of Physics and Technology, University of Jinan, Jinan, Shandong, 250022, People's Republic of China

[b] School of Physics, State Key laboratory of Crystal Materials, Shandong University, Jinan, Shandong, 250100, People's Republic of China

Group III-V films are of great importance for their potential application in spintronics and quantum computing. Search for two-dimensional III-V films with a nontrivial large-gap are quite crucial for the realization of dissipationless transport edge channels using quantum spin Hall (QSH) effects. Here we use first-principles calculations to predict a class of large-gap QSH insulators in functionalized TlSb monolayers ($TlSbX_2$; (X = H, F, Cl, Br, I)), with sizable bulk gaps as large as 0.22~0.40 eV. The QSH state is identified by $Z_2$ topological invariant together with helical edge states induced by spin-orbit coupling (SOC). Noticeably, the inverted band gap in the nontrivial states can be effectively tuned by the electric field and strain. Additionally, these films on BN substrate also maintain a nontrivial QSH state, which harbors a Dirac cone lying within the band gap. These findings may shed new light in future design and fabrication of QSH insulators based on two-dimensional honeycomb lattices in spintronics.

* Correspondence and requests for materials should be addressed to: zhchwsd@163.com

## I. Introduction

One of the grand challenges in condensed matter physics and material science is to develop room-temperature electron conduction without dissipation. Two-dimensional (2D) topological insulators (TIs), namely quantum spin Hall (QSH) insulators, are new states of quantum matter with an insulating bulk and metallic edge states[1-5]. Their helical edge states are spin-locked due to the protection of time-reversal symmetry (TRS), namely the propagation direction of surface electrons is robustly linked to their spin orientation[6], leading to dissipationless transport edge channels. However, the working temperature of QSH insulators in experiments like HgTe/CdTe[7, 8] and InAs/GaSb films [9, 10] are very low (below 10 K), limited by their small energy gap. The search of QSH insulators with large-gap is urgently required.

Chemical functionalization of 2D materials is an effective way to realize QSH state with desirable large-gaps. The most reported cases include hydrogenated or halogenated germanene and stanene [11-14]. These films are QSH insulators with gaps of 0.2~0.3eV, sufficient for practical applications at room temperature. Group V films, including Bi[15, 16] and Sb[17], are large-gap QSH insulators, when functionalized with hydrogen or halogens. Recently, the organic molecule ethynyl-functionalized stanene [18], and methyl-functionalized Bi and Sb films [19] have been reported to be good QSH insulators in the literature. Progress also undergoes simultaneously in experiments, Bi (111) film has been successfully grown on $Bi_2Te_3$ or $Bi_2Se_3$ substrates[20-23]. The common feature of these materials is that they all own 2D honeycomb-like crystal structures, indicating that 2D hexagonal lattice could be an excellent cradle to breed QSH insulators with SOC. These large-gap QSH insulators are essential for realizing many exotic phenomena and for fabricating new quantum devices that can operate at room temperature.

Group III-V materials are of importance applicable to semiconductor devices in semiconductor industry. Especially, the $\pi$ bonding between $p_z$ orbitals on group-III and V atoms can generally open a bulk gap with SOC, similar to graphene.[1] Different from the inversion-symmetry (IS) in graphene, the geometry of group III-V films is inversion-asymmetry (IAS) due to IS breaking. The previous works have shown that the IAS materials host many nontrivial phenomena such as the crystalline-surface-dependent topological electronic states,[24, 25] pyroelectricity,[26] topological *p*-*n* junctions,[27] as well as topological superconductivity,[28, 29] *et al*. However, one important characteristic of III-V films is that its unsaturated $p_z$ orbital is chemically active, due to the weak π-π interaction as caused by the bond length between III-V atoms (~3 Å). This feature, together with the out-of-plane orientation of $p_z$ orbital, facilitates strong orbital interaction with external environments, and thus its electronic properties are easily affected by adsorbates and substrates, unfavorable for practical applications in spintroncis.

As a representative, here we provide a systematical study on structural and topological properties of 2D TlSb monolayers functionalized with hydrogen and halogens, namely $TlSbX_2$ (X = H, F, Cl, Br, I). We find that the surface functionalization on TlSb, *i.e.*, saturating the $p_z$ orbital composed of TISb with hydrogen or halogens, can stabilize the 2D TlSb, according to the calculated phonon spectrum of $TlSbX_2$ films. All the systems are found to be QSH insulators, with the bulk gap in the range of 0.22~0.40 eV, tunable by external strain and electric field. A single pair of topologically protected helical edge states is established for these systems with the Dirac point locating in the bulk gap, and the odd numbers of crossings between edge states and Fermi level prove the nontrivial nature of these $TlSbX_2$ films. These findings may provide a new platform to design large-gap QSH insulator based on group III-V films, which is important for device application in spintronics.

## II. Computational method and details

To study the structural and electronic properties of $TlSbX_2$ (X = H, F, Cl, Br, I) films, our calculations were performed using the plane-wave basis Vienna *ab initio* simulation package known as VASP code.[30,31] We used the generalized gradient approximation (GGA) for the exchange and correlation potential, as proposed by Perdew-Burk-Ernzerhof (PBE),[32] the projector augmented wave potential (PAW)[33] to treat the ion-electron interactions. The energy cutoff of the plane waves was set to 500 eV with the energy precision of $10^{-6}$ eV. The Brillouin zone was sampled by using a 21×21×1 Gamma-centered Monkhorst-Pack grid. The vacuum space was set to 20 Å to minimize artificial interactions between neighboring slabs. SOC was included by a second vibrational procedure on a fully self-consistent basis. The phonon spectra were calculated using a supercell approach within the PHONON code.[34]

## III. Results and discussion

The geometric structure of $TlSbX_2$ (X = H, F, Cl, Br, I) are displayed in Fig. 1(a), in which the Tl or Sb atoms are saturated with X atoms on both sides of the plane in an alternating manner along the hexagonal axis, and thus breaks IS of $TlSbX_2$. Table 1 lists the calculated equilibrium lattice constants, buckling heights, as well as Tl-Sb, Tl-X, and Sb-X bond lengths after structural optimization. In comparison to pristine TlSb, the TI-Sb bonds in $TlSbX_2$ slightly expand, while the buckling changes differently due to the weakly hybridization between π and σ orbitals, stabilizing these structures. The stability of functionalized $TlSbX_2$ is studied through the formation energy defined as

$$E_f = \frac{E_{\mathrm{TlSbX2}} - (E_{\mathrm{TlSb}} + N_X E_{\mathrm{X}})}{N_X} \qquad (1)$$

where $E_{\mathrm{TlSbX2}}$ , $E_{\mathrm{TISb},}$ and $E_{\mathrm{X}}$ are the total energy of double-side functionalized $TlSbX_2$, pristine TlSb, and molecule $X_2$, respectively. $N_{\mathrm{X}}$ is the number of X atoms. The calculated formation energies for $TlSbH_2$, $TlSbF_2$, $TlSbCl_2$, $TlSbBr_2$, and $TlSbI_2$, are -1.862, -2.997, -1.613, -1.567, and -1.420 eV, respectively, suggesting that hydrogen or halogens are chemically bonded to TlSb, indicating a higher thermodynamic stability relative to their elemental reservoirs. The dynamic stability of $TlSbH_2$, as an example, is further confirmed by the phonon dispersion curves in Fig. 1(b), in which all branches have positive frequencies and no imaginary phonon modes, confirming the stability of $TlSbH_2$.

Figures 2(a) and (d) display the calculated band structure for $TlSbH_2$ and $TlSbF_2$ as representative examples, in which the red and blue lines correspond to band structures without and with SOC. In the absence of SOC, they are both gapless semimetal with the valence band maximum (VBM) and conduction band minimum (CBM) degenerate at the Fermi level. When takes SOC into account, the band structures of $TlSbH_2$ and $TlSbF_2$ produce a semimetal-to-semiconductor transition, with sizeable bulk-gaps of 0.22 eV and 0.40 eV, respectively. As observed in previously reported 2D TIs like $ZeTe_5$,[35] $HfTe_5$,[36] and GaSe,[37] graphene-like materials,[38-40] the SOC-induced band-gap opening at the Fermi level is a strong indication of the existence of topologically nontrivial phases.

An important character of the QSH insulator is helical edge states which is key to spintronic applications due to the ability to conduct dissipationless currents. Thus, we calculate the topological edge states by the Wannier90 package.[41] We construct the maximally localized Wannier functions (MLWFs) and fit a tight-binding Hamiltonian with these functions. The calculated edge Green's function[42] of semi-infinite $TlSbX_2$ (X = H, F) is shown in Figs. 3(a) and 3(d). One can see that all the edge bands connect completely the conduction and valence bands and span 2D bulk band gap, yielding a 1D gapless edge states. Besides, the counter-propagating edge states exhibit opposite spin-polarization, in accordance with the spin-momentum locking of 1D helical electrons. In addition, the Dirac point located at the band gap are calculated to have a high velocity of ~ $2.0 \times 10^5$ m/s, comparable to that of $5.5 \times 10^5$ m/s in HgTe/CdTe quantum well.[7, 8] All these consistently indicate that $TlSbX_2$ (X = H, F) are ideal 2D TIs. The topological states can be further confirmed by calculating topological invariant $Z_2$ after the band inversion. Due to IS breaking in $TlSbX_2$, the method proposed by Fu and Kane [43] cannot be used to calculate the $Z_2$ invariant. Thus, a method independent of the presence of IS is needed. Here, we introduce the evolution of Wannier Center of Charges (WCCs) [44] to calculate the $Z_2$ invariant, which can be expressed as:

$$Z_2 = P_\theta(T/2) - P_\theta(0) \tag{2}$$

which indicates the change of time-reversal polarization ($P_\theta$) between the 0 and *T*/2. Then the WFs related with lattice vector R can be written as:

$$|R,n\rangle = \frac{1}{2\pi}\int_{-\pi}^{\pi} dk e^{-ik(R-x)} |u_{nk}\rangle \tag{3}$$

Here, a WCC $\bar{x}_n$ can be defined as the mean value of $\langle 0n|\hat{X}|0n\rangle$, where the $\hat{X}$ is the position operator and $|0n\rangle$ is the state corresponding to a WF in the cell with $R$ = 0. Then we can obtain

$$\bar{x}_n = \frac{i}{2\pi}\int_{-\pi}^{\pi} dk \langle u_{nk}|\partial_k|u_{nk}\rangle \tag{4}$$

Assuming that $\sum_\alpha \bar{x}_\alpha^S = \frac{1}{2\pi}\int_{BZ} A^S$ with $S = I$ or $II$, where summation in $\alpha$ represents the occupied states and $A$ is the Berry connection. So we have the format of $Z_2$ invariant:

$$Z_2 = \sum_\alpha [\bar{x}_\alpha^I(T/2) - \bar{x}_\alpha^{II}(T/2)] - \sum_\alpha [\bar{x}_\alpha^I(0) - \bar{x}_\alpha^{II}(0)] \tag{5}$$

The $Z_2$ invariant can be obtained by counting the even or odd number of crossings of any arbitrary horizontal reference line. In Figs. 3(c) and 3(f), we display the evolution lines of Wannier function centers (WFC) for $TlSbH_2$ and $TlSbF_2$, respectively. It can be seen that the WFC evolution curves cross any arbitrary reference lines odd times, thus yielding $Z_2 = 1$.

Now, we turn to the physics of QSH effect in $TlSbX_2$. Since the decorated atoms hybridizes strongly with the dangling bonds of $p_z$ orbital in TISb, it effectively removes the $p_z$ bands away from the Fermi level, leaving only the *s* and $p_{x,y}$ orbitals, as displayed in Figs. 2(b-c) and 2(e-f). However, through projecting the bands onto different atomic orbitals, we find that there are two scenarios for the effect of SOC on the bands around the Fermi level, in which the *s* and $p_{x,y}$ band inversion are different from each other. For $TlSbH_2$, at the Γ point, the two $p_{x,y}$ orbitals from TI and Sb atoms are energy degenerate, while the bands away from the Γ point are well separated due to orbital splitting. The Fermi level is located between one *s* and two $p_{x,y}$ orbitals, rendering the *s* above $p_{x,y}$ orbitals in energy, thus forming a normal band order, similar to the cases in conventional III-V semiconductors. While for $TlSbF_2$, the band structures are changed drastically, as shown in Figs. 2(d-f). In sharp contrast to $TlSbH_2$, the band order at the Γ point is inverted, *i.e.*, the *s* is shifted below two $p_{x,y}$ orbitals. These two different band orders may be attributed to the chemical bonding and orbital splitting between TI and Sb atoms. To further understand the physics of band inversion, we display in Fig. 4 the orbital evolution at the Γ point around the Fermi level of $TlSbH_2$ and $TlSbF_2$ films. On can see that, the chemical bonding and crystal field splitting between TI and Sb atoms make the *s* and $p_{x,y}$ orbital split into the bonding and anti-bonding states, *i.e.*, $s^{\pm}$ and $p^{\pm}_{x,y}$, which the superscripts + and − represent the parities of corresponding states, respectively. As displayed in Fig. 4(a), the $s^{+}$ orbital for hydrogenated one is significantly higher somewhere above $p^{-}_{x,y}$ orbital of Tl and Sb atoms under the effect of crystal field. The inclusion of SOC makes the degeneracy of $p^{-}_{x,y}$ orbital split into $p^{-}_{x+iy,\uparrow}$ & $p^{-}_{x-iy,\downarrow}$ and $p^{-}_{x-iy,\uparrow}$ & $p^{-}_{x+iy,\downarrow}$, leading $s^{+}$ locate in between them. On the other hand, for $TlSbF_2$ in Fig. 4(b), the larger lattice constant results in a weaker *s-p* hybridization, and accordingly a smaller energy separation between the bonding and anti-bonding states. Thus, the $s^{-}$ orbital is downshifted while the $p^{+}_{x,y}$ is upshifted, *i.e.*, the $s^{-}$ will be occupied, while the degenerate $P^{+}_{x,y}$ is half occupied, resulting in semi-metallic character (Fig. 2(d)).Though the inclusion of SOC make also the degeneracy of $p^{+}_{x,y}$ orbital split into $p^{+}_{x+iy,\uparrow}$ & $p^{+}_{x-iy,\downarrow}$ and $p^{+}_{x-iy,\uparrow}$ & $p^{+}_{x+iy,\downarrow}$, but its *s-p* band order are not changed. As a result, the mechanism of QSH effect can be roughly classified into two categories: *i.e.*, type-I: SOC-induced *p-s-p* TI ($TlSbH_2$), and type-II: Chemical bonding induced *p-p-s* TI ($TlSbF_2$). Obviously, it is the *s* orbital insertion into $p^{+}_{x+iy,\uparrow}$ & $p^{+}_{x-iy,\downarrow}$ and $p^{+}_{x-iy,\uparrow}$ & $p^{+}_{x+iy,\downarrow}$ that the topological bulk-gap (0.22 eV) of $TlSbH_2$ is smaller twice than that (0.40 eV) of $TlSbF_2$ film.

Here, we wish to point out that fluorination in TlSb is not the only way to achieve large-gap QSH state, the same results can be obtained by decorating the surface with otherwise halogens, such as Cl, Br, and I. We thus performed calculations for $TlSbX_2$ (X = Cl, Br, I) films to check their topological properties, as illustrated in Figs. S1. Table 1 summarizes their lattice constants, Sb-X and Tl-X bond lengths, and nontrivial QSH bulk-gaps at their equilibrium states. The results demonstrate that the electronic structures of all these $TlSbX_2$ films are similar to $TlSbF_2$, and exhibit nontrivial topological states (Fig. S2). Interestingly, as can be seen in Fig. 4(c) and Fig. S1, the global gaps in QSH state are obtained to be 0.34, 0.32, and 0.29 eV for $TlSbCl_2$, $TlSbBr_2$ and $TlSbI_2$, respectively, which are sufficient for practical applications at room temperature. However, when comparing the band gaps with each other, we can find some fascinating phenomena that the global band gaps of these systems monotonically decrease in the contrary order of $TlSbF_2 > TlSbCl_2 > TlSbBr_2 > TlSbI_2$. It is known that, from F to I, the SOC becomes stronger in the order of $F < Cl < Br < I$, thus the SOC-induced bulk-gap should be increased correspondingly. This interesting contradiction can be attributed to the variation of band components of Tl and Sb atoms near the Fermi level, as the band splitting driven by SOC can directly determine QSH gap. As shown in Fig. 4(c), the ratio from the Sb-$p_{x,y}$ to X-$p_{x,y}$ orbital at Γ point decreases in the order of $TlSbF_2 > TlSbCl_2 > TlSbBr_2 > TlSbI_2$. Similar results are obtained for the ratio of Tl-$p_{x,y}$ to X-$p_{x,y}$ component in Fig. 4(d). Considering that the Tl and Sb atoms exhibits stronger SOC strength than halogens, it is expected that the larger the ratio is, the larger the contribution to the states near the Fermi level, and thus the larger the SOC strength.

Strain engineering is a powerful approach to modulate electronic properties and topological natures in 2D materials, and thus it is interesting to study these effects in $TlSbX_2$ films. We employ an external strain on these monolayers maintaining the crystal symmetry by changing its lattices as $\varepsilon = (a\text{-}a_0)/a_0$, where $a$ ($a_0$) is the strained (equilibrium) lattice constants. As shown in Figs. 5(a) and (b), the magnitude of nontrivial bulk-gaps of $TlSbH_2$ and $TlSbF_2$ can be modified significantly by strain, implying the interatomic coupling can modulate the topological natures of these systems. For $TlSbH_2$, with increasing the strain, the CBM is continuously to shift downward to the Fermi level, while the VBM increases reversibly, leading the band gap to decrease significantly (Fig. 5(a)). When the critical value reaches up to -3.8 %, a semi-metallic state with zero density of states at the Fermi level occurs. If increases the strain beyond -3.8 %, a trivial topological phase appears. While for $TlSbX_2$ (see also Fig. 5(b) and Fig. S3), both the direct and indirect band gaps decreases steadily with respect to tensile strain. Especially, these QSH states are robust with the strain in the range of -8 ~ 16 %. Such robust topology against lattice deformation makes $TlSbX_2$ easier for experimental realization and characterization on different substrate.

On the other hand, from the perspective of potential device applications, the ability to control topological electronic properties via the vertical electric field (*E*-field) is highly desirable. Thus, we study the change of band gaps of $TlSbH_2$ and $TlSbF_2$ under different vertical *E*-field, as shown in Figs. 5(c) and (d). One can see that the band gaps increase monotonically with increasing *E*-field strength for both cases. For $TlSbF_2$ (Fig. 5(d)), under -0.8 V $Å^{-1}$≤*E*-field ≤0.8 V $Å^{-1}$, the trend of band gaps increase monotonically from 0.34 eV to 0.41 eV, with $E_\Gamma$ larger than $E_g$ significantly. While for $TlSbH_2$, when *E*-field ≤ -5.5 V $Å^{-1}$, it is a normal insulator. But for large *E*-field (> -0.4 V $Å^{-1}$), it becomes a QSH insulator, along with $E_\Gamma$ being almost equal to $E_g$. Noticeably, if *E*-field is in the range of ± 8%, the nontrivial bulk-gaps of other $TlSbX_2$ (X = Cl, Br, I) are still very large (~0.2−0.5 eV) (Fig. S3), allowing for viable applications at room temperature. The predicted QSH insulators tuned by vertical *E*-field may provide a platform for realizing topological field-effect transistor (TFET).

The substrate materials are inevitable in device application, thus a free-standing film must eventually be deposited or grown on a substrate. As a 2D large-gap insulator with a high dielectric constant, the BN sheet has been successfully used as the substrate to grow graphene or assemble 2D stacked nanodevices[45, 46]. Thus, we use it as a substrate to support $TlSbX_2$ films. Figs. 6(a-b) show the geometrical structures of $TlSbH_2$ and $TlSbF_2$ on (2×2) BN sheet, where the lattice mismatch is only about 1.68% and 2.80%, respectively. After full relaxation with van der Waals (vdW) forces[47], they almost retain the original structure with a distance between the adjacent layers of 3.35 Å. The calculated binding energy is about -69, -87, -98, -108, and -114 meV for $TlSbH_2$, $TlSbF_2$, $TlSbCl_2$, $TlSbBr_2$, and $TlSbI_2$ per unit cell, respectively, showing that they are typical van der Waals heterostructures. The calculated band structure with SOC is shown in Figs. 6(c) and (d). In these weakly coupled systems, $TlSbH_2$ on the BN sheet remains semiconducting, there is essentially no charge transfer between the adjacent layers, and the states around the Fermi level are dominantly contributed by $TlSbH_2$. If we compare the bands of $TlSbH_2$ with and without BN substrate, little difference is observed. Similar results are also found for all halogenated $TlSbX_2$ films on BN substrate (see also Fig. S4), except that $TlSbF_2$ on the BN sheet exhibits a metallic state. Evidently, all $TlSbX_2$ films on BN substrate are robust QSH insulators.

## IV. Conclusions

To conclude, on the basis of first-principles calculations, we predict a class of new QSH insulator of $TlSbX_2$ (X = H, F, Cl, Br, I) films, with a sizable bulk gap (0.22 ~ 0.40 eV), allowing for viable applications in spintronic devices. Two mechanisms, type-I: SOC-induced *p*-*s*-*p* type TI ($TlSbH_2$), and type-II: the chemical bonding induced *p*-*p*-*s* type TI (Halogenated ones) are obtained, significantly different from one in TISb monolayer. The topological characteristic of $TlSbX_2$ films are confirmed by the $Z_2$ topological order and topologically protected edge states. Furthermore, the band gap and topological phase transition could be tuned by the external strain and vertical *E*-field. When $TlSbX_2$ deposited on BN substrate, both the band gaps and low-energy electronic structures are only slightly affected by the interlayer coupling from the substrate. These predicted QSH insulators and their vdW heterostructures may provide a platform for realizing low-dissipation quantum electronics and spintronics devices.

---

**Acknowledgments**

This work was supported by the National Natural Science Foundation of China (Grant No. 11274143, 11434006, 61172028, and 11304121), and Research Fund for the Doctoral Program of University of Jinan (Grant no. XBS1433).

**Additional information**

Competing financial interests: The authors declare no competing financial interests.

**References**

1. M. Z. Hasan, C. L. Kane, *Rev. Mod. Phys.* 2010, **82**, 3045-3067.

2. X. L. Qi, S. C. Zhang, *Rev. Mod. Phys.* 2011, **83**, 1057-1110.

3. B. Yan, S. C. Zhang, *Rep. Prog. Phys.* 2012, **75**, 096501.

4. J. E. Moore, *Naturenanotechnology* 2013, **8**, 194-198.

5. B. Rasche, A. Isaeva, M. Ruck, S. Borisenko, V. Zabolotnyy, B. Buchner, K. Koepernik, C. Ortix, M. Richtern, J. van den Brink, *Nat. Mater.* 2013, **12**, 422-425.

6. H. J. Zhang, C. X. Liu, X. L. Qi, X. Dai, Z. Fang, S. C. Zhang, *Nat. Phys.* 2009, **5**, 438-442.

7. E. N. Lima, T. M. Schmidt, *Phys. Rev. B* 2015, **91**, 075432.

8. P. Dziawa, B. J. Kowalski, K. Dybko, *et al*. *Nat. mater.* 2012, **11**, 1023-1027.

9. L. Du, I. Knez, G. Sullivan, R. R. Du, *Phys. Rev. Lett.* 2015, **114**, 096802.

10. C. X. Liu, T. L. Hughes, X. L. Qi, K. Wang, S. C. Zhang, *Phys. Rev. Lett.* 2008, **100**, 236601.

11. C. C. Liu, J. Hua, Y. G. Yao, *Phys. Rev. B* 2011, **84**, 195430.

12. F. Yang, L. Miao, Z. F. Wang, M.Y. Yao, F. Zhu, Y. R. Song, *et al*. *Phys. Rev. Lett.* 2012, **109**, 016801.

13. S. C. Wu, G. C. Shan, B. H. Yan, *Phys. Rev. Lett.* 2014, **113**, 256401.

14. P. Z. Tang, P. C. Chen, W. D. Cao, *et al. Phys. Rev. B* 2014, **90**, 121408.

15. C. C. Liu, S. Guan, Z. G. Song, S. Y. Yang, J. B. Yang, Y. G. Yao, *Phys. Rev. B* 2014, **90**, 085431.

16. K. H. Jin, S. H. Jhi, *Sci. Rep.* 2015, **5**, 8426.

17. Z. G. Song, C. C. Liu, J. B. Yang, J. Z. Han, M. Ye, B. T. Fu, Y. C. Yang, Q. Niu, J. Lu, Y. G. Yao, *NPG Asia Mater.* 2014, **6**, e147.

18. R. W. Zhang, C. W. Zhang, W. X. Ji, *et al*. *New J. Phys.* 2015, **17**, 083036.

19. Y. D. Ma, Y. Dai, L. Z. Kou, *et al*. *Nano letters* 2015, **15**, 1083.

20. F. Yang, L. Miao, Z. Wang, *et al*. *Phys. Rev. Lett.* 2012, **109**, 016801.

21. T. Hirahara, N. Fukui, T. Shirasawa, *et al*. *Phys. Rev. Lett.* 2012, **109**, 227401.

22. N. Fukui, T. Hirahara, T. Shirasawa, T. Takahashi, K. Kobayashi, S. Hasegawa, *Phys. Rev. B* 2012, **85**, 115426.

23. Z. F. Wang, M. Y. Yao, W. M. Ming, F. F. Zhu, C. H. Liu, C. L. Gao, D. Qian, J. F. Jia, F. Liu, *Nat. Commun.* 2013, **4**, 1384.

24. S. Murakami, *Phys. Rev. Lett.* **97**, 236805 (2006).

25. M. S. Bahramy, B. J. Yang, R. Arita. N. Nagaosa, *Nature Comm.* 2012, **3**, 679.

26. X. Wan, A. M. Turner, A. Vishwanath, S. Y. Savrasov, *Phys. Rev. B* 2011, **83**, 205101.

27. J. Wang, X. Chen, B. F. Zhu, S. C. Zhang, *Phys. Rev. B* 2012, **85**, 235131.

28. E. Bauer, G. Hilscher, H. Michor, Ch. Paul, E. W. Scheidt, A. Gribanov, Y. Seropegin, H. Noel, M. Sigrist, P. Rogl, *Phys. Rev. Lett.* 2004, **92**, 027003.

29. P. A. Frigeri, D. F. Agterberg, A. Koga, M. Sigrist, *Phys. Rev. Lett.* 2004, **92**, 097001.

30. G. Kresse, J. Furthmüller, *Phys. Rev. B.* 1996, **54**, 11169.

31. G. Kresse, J. Furthmüller, *Comput. Mater. Sci.* 1996, **6**, 15-50.

32. J. P. Perdew, K. Burke, M. Ernzerhof, *Phys. Rev. Lett.* 1996, **77**, 3865.

33. P. E. Blöchl, *Phys. Rev. B.* 1994, **50**, 17953.

34. H. Zimmermann, R. C. Keller, P. Meisen, M. Seelmann-Eggebert, *Surf. Sci.* 1997, **904**, 377-379.

35. Q. Liu, X. Zhang, L. B. Abdalla, *et al. Nano letters* 2015, **15**, 1222-1228.

36. H. Weng, X. Dai, Z. Fang, *Phys. Rev. X*, 2014, **4**, 011002.

37. Z. Zhu, Y. Cheng, U. Schwingenschlo□gl, *Phys. Rev. Lett.* 2012, **108**, 266805.

38. C. L. Kane, E. J. Mele, *Phys. Rev. Lett.* 2005, **95**, 226801.

39. C. C. Liu, W. X. Feng, Y. G. Yao, *Phys. Rev. Lett.* 2011, **107**, 076802.

40. Y. Xu, B. H. Yan, H. J. Zhang, J. Wang, G. Xu, P. G. Tang, W. H. Duan, S. C. Zhang, *Phys. Rev. Lett.* 2013, **111**, 136804.

41. A. A. Mostofi, J. R. Yates, Y. S. Lee, I. Souza, D. Vanderbilt, N. Marzari, *Computer Physics Communications* 2008, **178**, 685.

42. M. P. L. Sancho, J. M. L. Sancho, J. Rubio, *Journal of Physics F: Metal Physics* 1984, **14**, 1205.

43. L. Fu, C. L. Kane, *Phys. Rev. B* 2009, **79**, 161408.

44. R. Yu, X. L. Qi, A. Bernevig, Z. Fang, X. Dai, *Phys. Rev. B.* 2011, **84**, 075119.

45. K. K. Kim, A. Hsu, X. Jia, *et al*. *Nano letters* 2011, **12**, 161-166.

46. W. Yang, G. Chen, Z. Shi, *et al*. *Nature materials* 2013, **12**, 792-797.

47. J. Klimeš, D. R. Bowler, A. Michaelides, *Phys. Rev. B* 2011, **83**, 195131.

**Table I** Calculated structural parameters of the $TlSbX_2$ (X = H, F, Cl, Br, I) films, including the lattice parameter $a$(Å), buckled height $h$(Å), bulk gap $\boldsymbol{E_g}$ (eV) and $\boldsymbol{E_\Gamma}$ (eV), while the $d_{TlSb}$, $d_{Tl\text{-}X}$, and $d_{Sb\text{-}X}$ are the bond lengths of Tl-Sb, Tl-X, and Sb-X atoms, respectively (in Å ).

| **Structure** | $\boldsymbol{a}$(Å) | $h$(Å) | $\boldsymbol{E_g}$(eV) | $\boldsymbol{E_\Gamma}$(eV) | $d_{TlSb}$(Å) | $d_{Tl\text{-}X}$(Å) | $d_{Sb\text{-}X}$(Å) |
|---|---|---|---|---|---|---|---|
| *TlSb* | 4.810 | 0.78 | 0.28 | 0.29 | 2.88 | — | — |
| $TlSbH_2$ | 4.911 | 0.87 | 0.22 | 0.25 | 2.97 | 2.13 | 1.78 |
| $TlSbF_2$ | 5.270 | 0.42 | 0.40 | 0.47 | 3.07 | 2.15 | 1.95 |
| $TlSbCl_2$ | 5.268 | 0.59 | 0.34 | 0.42 | 3.06 | 2.51 | 2.37 |
| $TlSbBr_2$ | 5.098 | 0.70 | 0.32 | 0.44 | 3.03 | 2.63 | 2.53 |
| $TlSbI_2$ | 5.050 | 0.78 | 0.29 | 0.49 | 3.02 | 2.82 | 2.74 |

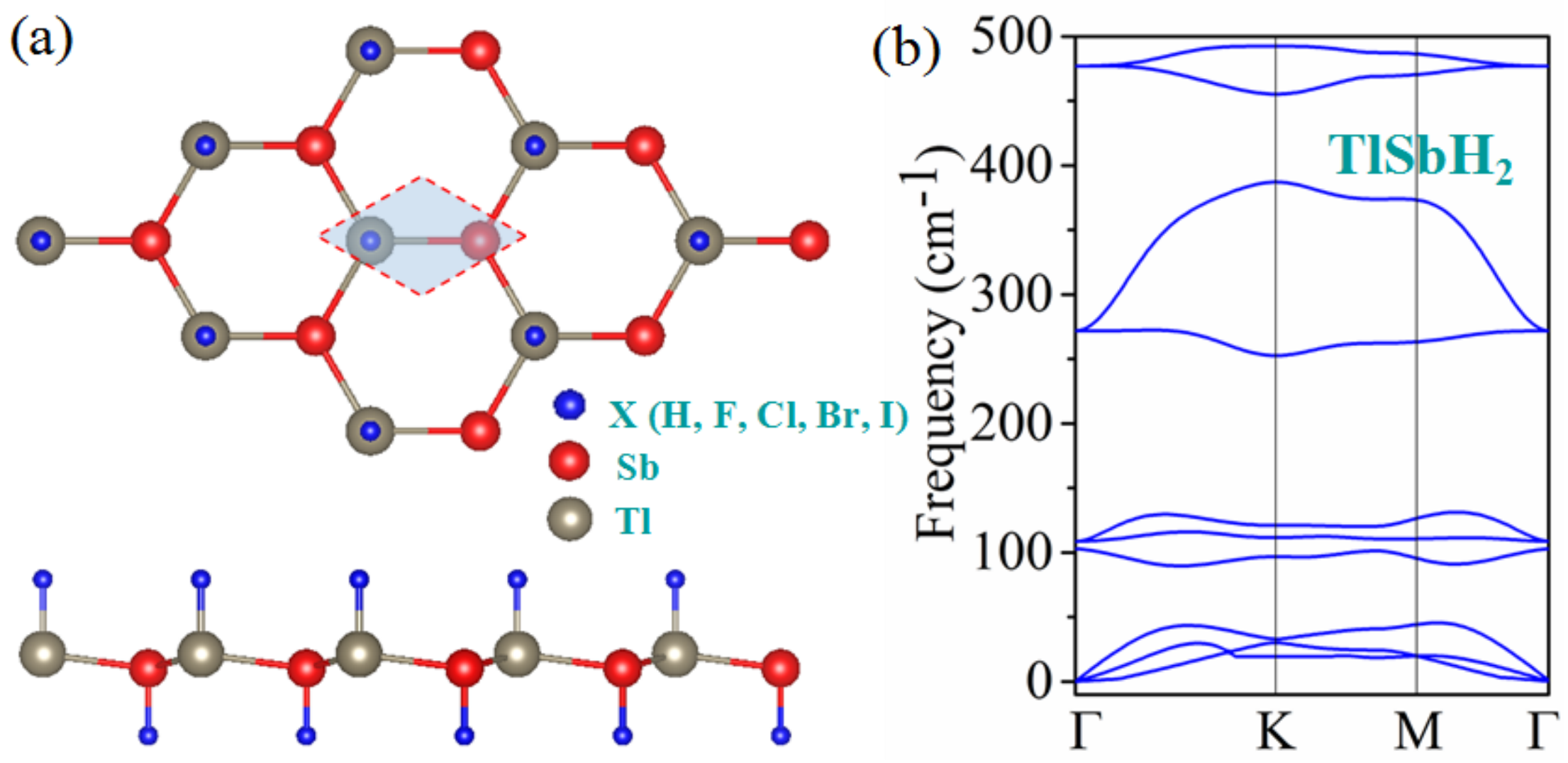

**Figure 1** (a) Top and side views of the geometrical structures of $TlSbX_2$ (X=H, F, Cl, Br, I). Blue, red, and gray balls denote hydrogen & halogen, Sb, and Tl atoms, respectively. Shadow area in (a) presents a unit cell. (b) Phonon band dispersion for $TlSbH_2$.

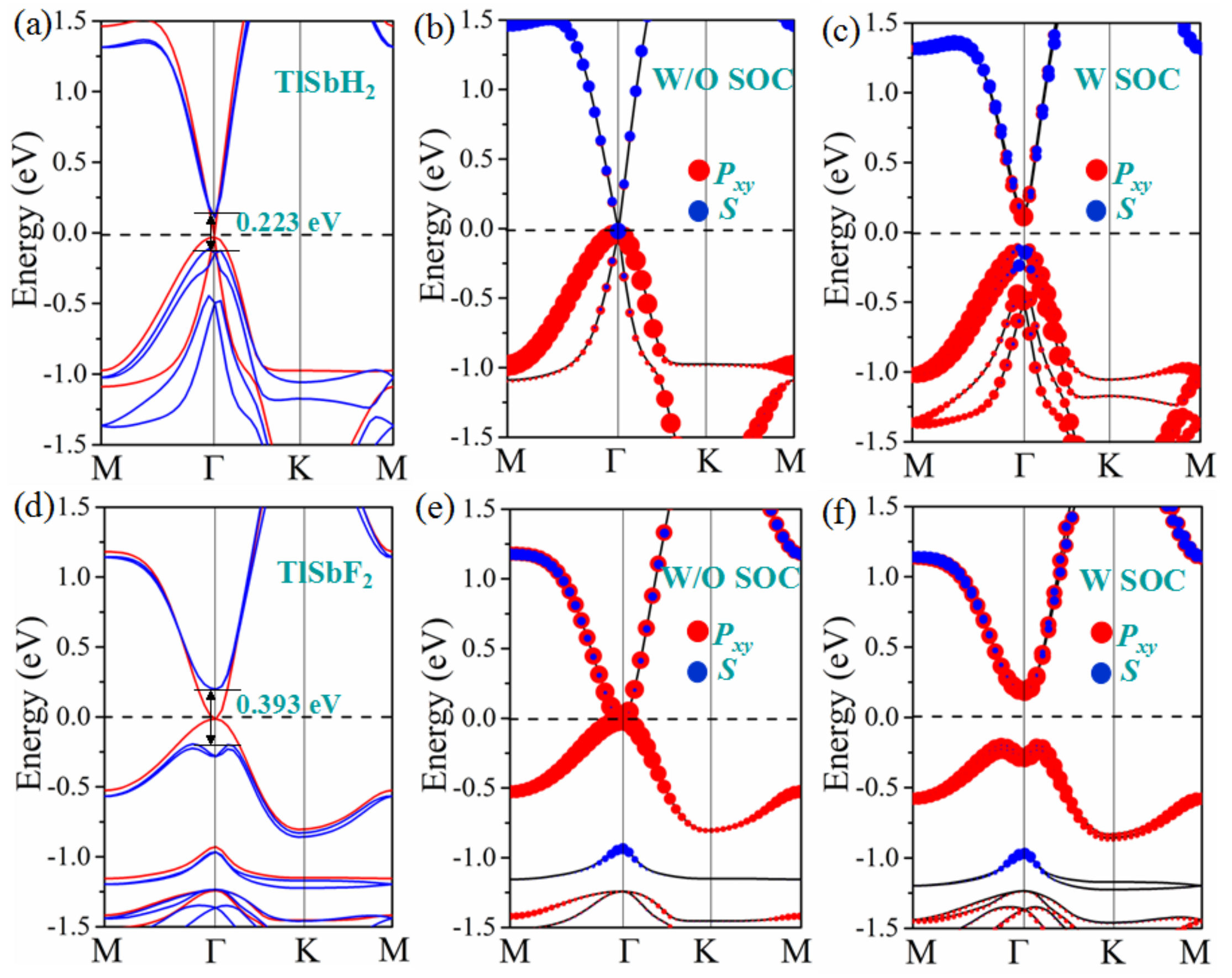

**Figure 2** The calculated band structures for (a) $TlSbH_2$ and (d) $TlSbF_2$ with and without SOC. The red lines correspond to band structures without SOC, and the blue lines correspond to band structures with SOC. (b) and (c) Orbital-resolved band structures of $TlSbH_2$, as well as (e) and (f) $TlSbF_2$, respectively. The blue dots represent the contributions from the *s* atomic orbital, and the red dots represent contributions from the $p_{x,y}$ atomic orbitals of Tl and Sb atoms.

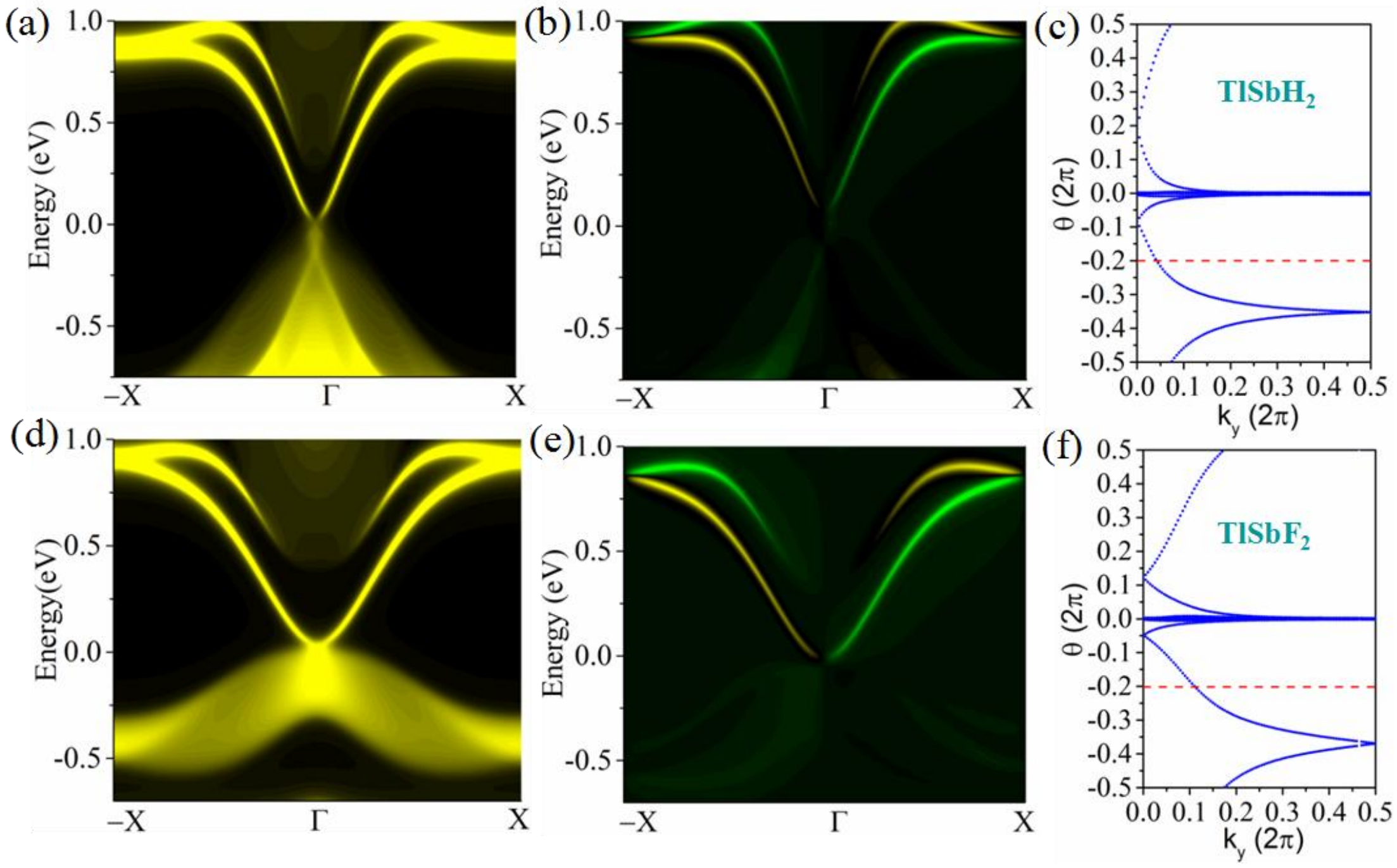

**Figure 3** Total (left panel) and spin (right panel) edge density of states for (a) and (b) $TlSbH_2$, (d) and (e) $TlSbF_2$. In the spin edge plot, yellow/green lines denote the spin up/down polarization. Evolutions of Wannier centers along $k_y$ are presented in (c) $TlSbH_2$ and (f) $TlSbF_2$. The evolution lines (blue dot lines) cross the arbitrary reference line (red dash line parallel to $k_y$) with an odd number of times, thus yielding $Z_2 = 1$.

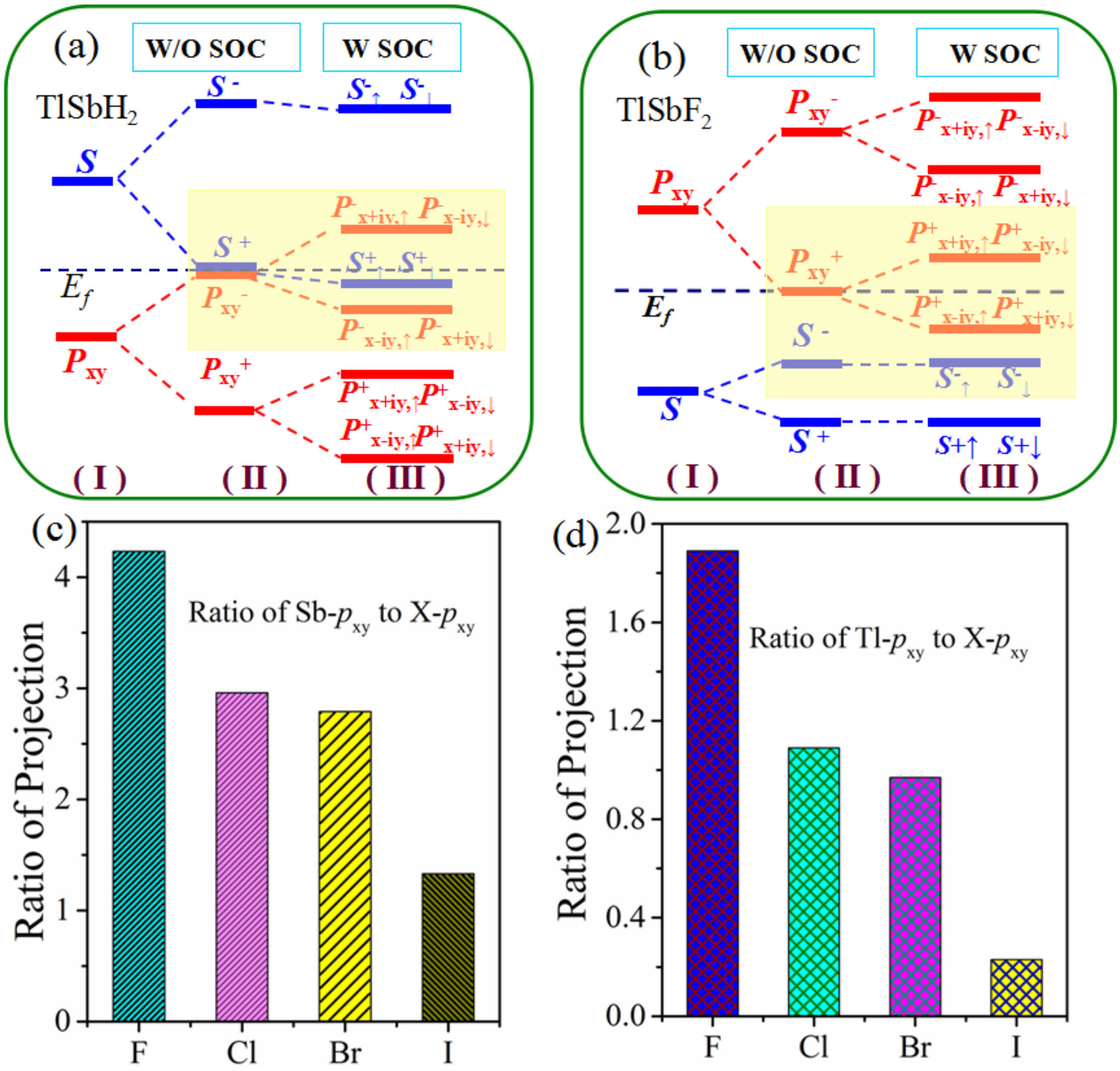

**Figure 4** The evolution of atomic *s* and $p_{x,y}$ orbitals without SOC and with SOC of (a) $TlSbH_2$ and (b) $TlSbF_2$. The horizontal blue dashed lines indicate the Fermi level. (c) The ratio of Sb-$p_{x,y}$ to X-$p_{x,y}$ component in the $p_{x,y}$ orbital at the Fermi level. (d) The ratio of TI-$p_{x,y}$ to X-$p_{x,y}$ component in the $p_{x,y}$ orbital at the Fermi level.

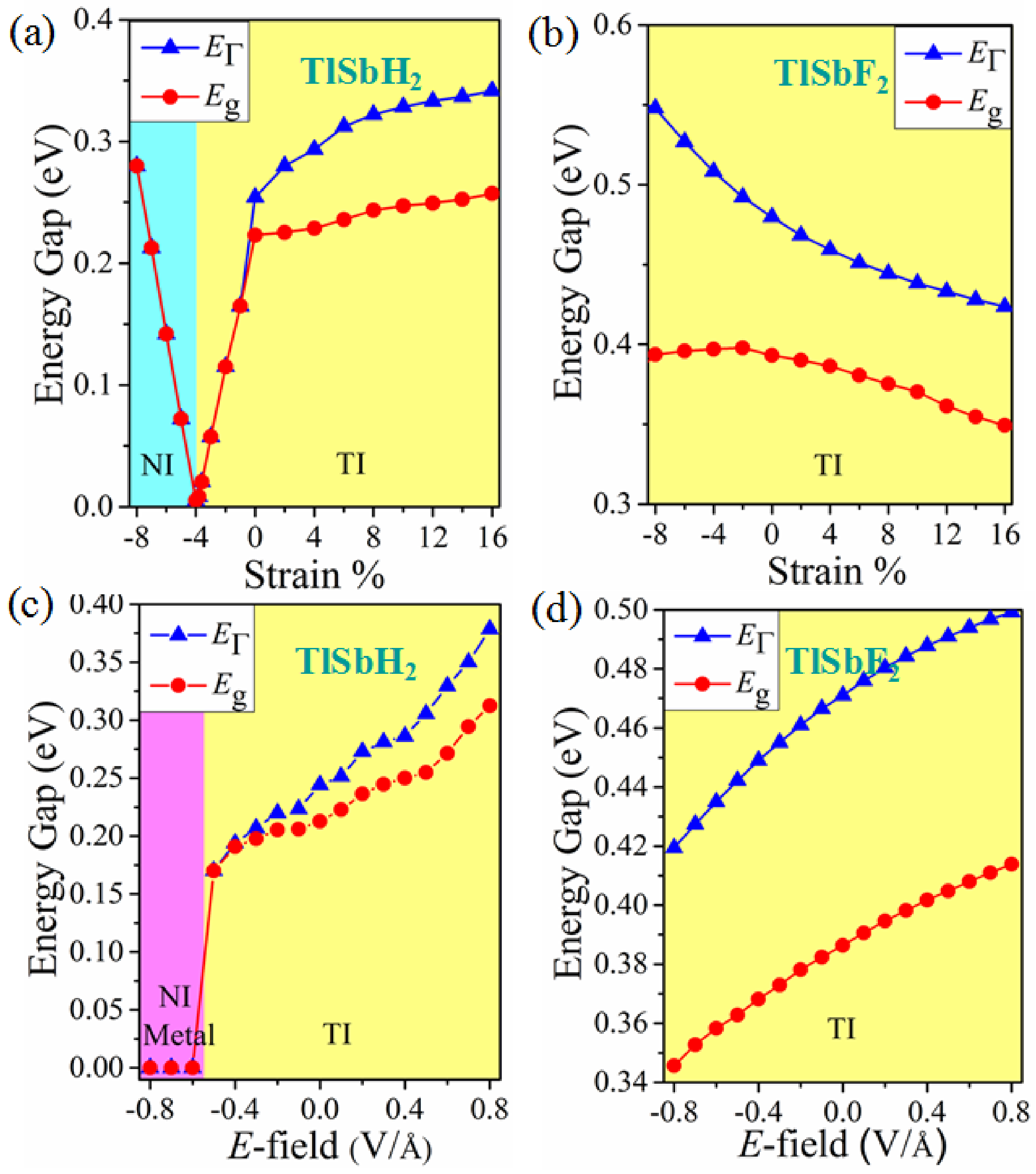

**Figure 5** The dependence of band gap of the strain (a) $TlSbH_2$ and (b) $TlSbF_2$, and the electric field (c) $TlSbH_2$ and (d) $TlSbF_2$, respectively.

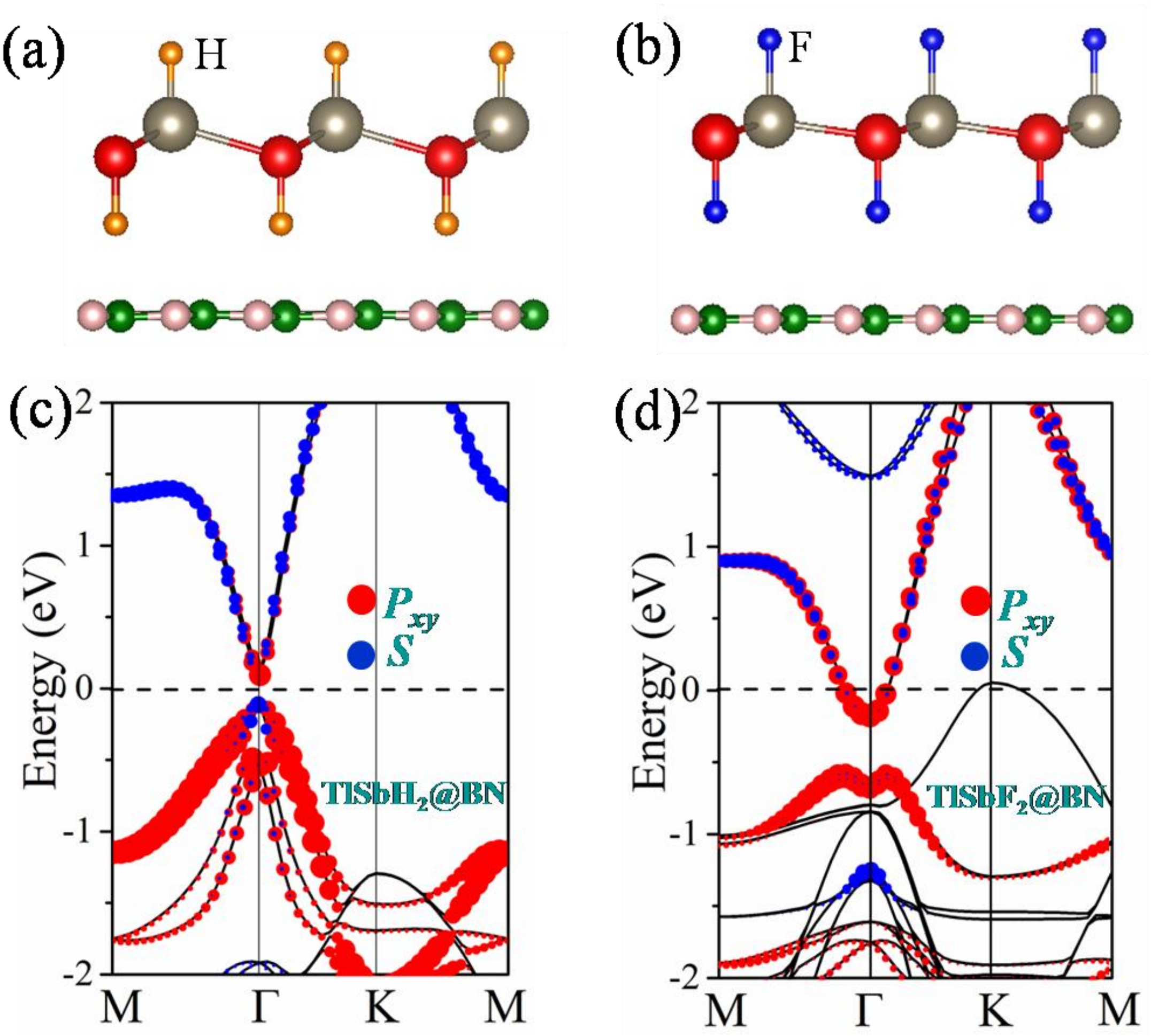

**Figure 6** Crystal structures of $TlSbX_2$ grown on BN substrate from the top and side view for (a) $TlSbH_2$ and (b) $TlSbF_2$. (c) $TlSbH_2$ and (d) $TlSbF_2$ correspond to the orbital-resolved band structures with SOC.